\documentstyle[namedreferences]{kluwer}

\newdimen\thicksize
\newdimen\thinsize
\def\th{\thinspace}

\def\qquad{\quad\quad}
\def\ngth{\negthinspace}

\def\frac#1#2{{\textstyle{ #1 \over #2}}}

\def\ahalf {{\scriptstyle {1\over 2}}}

\def\eg{{{\it e.g.}\ }}
\def\etal{{\it et al.\ }}

\def\cf{{\it cf.\ }}

\def\ie{{{\it i.e.}\ }}

\def\viz{{\it viz.\ }}
\def\vs{{\it vs.\ }}

\def\at{{\rm\char'100}}
\def\ni{\noindent}

\def\Teff{{$T_{e\!f\!f} $}}
\def\Mo{{$M_\odot $}}
\def\Lo{{$L_\odot $}}

\def\lgt{ \raise4pt \hbox{$<$}\kern-9pt\lower1.5pt \hbox{$>$}}
\def\glt{\raise4pt \hbox{$<$}\kern-9pt\lower1.5pt\hbox{$>$}}

\def\approxgt{\raise4pt \hbox{$>$}\kern-9pt\lower1.5pt\hbox{$\sim$}}
\def\approxlt{\raise4pt \hbox{$<$}\kern-9pt\lower1.5pt\hbox{$\sim$}}

\def\ihp   {{i+\scriptscriptstyle{1\over 2} }}
\def\ihpp   {{i+\scriptscriptstyle{3\over 2} }}
\def\ihm   {{i-\scriptscriptstyle{1\over 2} }}

\def\nhp   {{n+\scriptscriptstyle{1\over 2} }}
\def\nhm  {{n-\scriptscriptstyle{1\over 2} }}

\def\fpi3  {{{4\pi \over 3}}}
\def\CS   {{{\cal S}}}

\runningtitle{An Adaptive Code}
\runningauthor{Buchler et al.}

 \begin{opening}

\title{AN ADAPTIVE CODE FOR RADIAL STELLAR MODEL PULSATIONS}

\author{
J. ROBERT \surname{BUCHLER}\thanks
{e--mail: buchler\at phys.ufl.edu},
ZOLT\'AN \surname{KOLL\'ATH}\thanks
{on leave from Konkoly Observatory, Budapest, Hungary}
and ARIEL \surname{MAROM}
}
\institute{
Physics Department, University of Florida, Gainesville, FL 32611
}

 \end{opening}

\begin{document}

 \begin{abstract}

We describe an implicit 1--D adaptive mesh hydrodynamics code that is specially
tailored for radial stellar pulsations.  In the Lagrangean limit the code
reduces to the well tested Fraley scheme.  The code has the useful feature that
unwanted, long lasting transients can be avoided by smoothly switching on the
adaptive mesh features starting from the Lagrangean code.  Thus, a limit cycle
pulsation that can readily be computed with the relaxation method of
Stellingwerf will converge in a few tens of pulsation cycles when put into the
adaptive mesh code.  The code has been checked with two shock problems, \viz
Noh and Sedov, for which analytical solutions are known, and it has been found
to be both accurate and stable.  Superior results were obtained through the
solution of the total energy (gravitational + kinetic + internal) equation
rather than that of the internal energy only.

 \end{abstract}


\input psfig
\def\rahmen#1{}

\section{Introduction}

The numerical hydrodynamical modelling of radial stellar pulsations goes back
to the pioneering efforts of Christy (1966) in the mid 1960s.  Reviews of the
stellar pulsation problem can be found in the literature (\eg Cox \& Giuli
1969, Cox 1980, Buchler 1990).  In our own past numerical work we have adopted
the hydrodynamics code developed by Stellingwerf (1974, 1975) on the basis of
Fraley's (1968) implicit Lagrangean scheme.  This is a very stable and robust
code, and in its relaxation version it is an excellent tool for computing
periodic pulsations and their linear stability (Floquet analysis).  The use of
the code has produced good results for a variety of stellar models ranging from
RR Lyrae (\eg Stellingwerf 1975, Kov\'acs \& Buchler Kov\'acs 1988a) and
Cepheids (\eg Moskalik, Buchler \& Marom 1992) to irregularly pulsating
variables (Kov\'acs \& Buchler 1988b), and good overall agreement with
observations can be achieved.  As these theoretical studies have progressed,
and as more and better observational has accumulated it has become increasingly
clear that a better code is needed (\eg Buchler 1990) if further advances are
to be made (Kov\'acs 1990, Kov\'acs \& Buchler 1993).

In fact, right from the beginning of Cepheid modelling misgivings were
expressed about the spatial resolution that can be obtained with a Lagrangean
grid.  The earliest adaptive attempts made to abate this deficiency were those
of Castor, Davis \& Davison (1977) and Aikawa \& Simon (1983), Simon \& Aikawa
(1986), but these codes were not energy conserving.  More recently, taking
advantage of progress in numerical mathematics, a couple of efforts
specifically aimed at stellar pulsations, and parallel to ours have been made
(Dorfi \& Drury 1987, Dorfi \& Feuchtinger 1991, Feuchtinger \& Dorfi 1994,
Cox, Deupree \& Gehmeyr 1992, Gehmeyr 1992a,b, 1993).  The three codes differ
in many respects, however.  In particular, the other two codes solve the
internal energy equation whereas we solve the total energy equation (While
analytically these equations are equivalent, the tests that we describe below
show that, numerically, the solution of the total energy equation is more
accurate.  Furthermore our grid equation while similar to Dorfi's uses a mass
density rather than a radius density.  A comparison of the three codes on a few
standard stellar models would be desirable.

Our strategy in developing an adaptive code has been very conservative.
Because of the success of the Stellingwerf-Fraley scheme we have decided to
build an adaptive code on top of this Lagrangean scheme.  By putting the mass
and total energy equations in conservative form, and by properly defining the
fluxes, we can write the adaptive scheme so that, {\sl globally}, mass and
total energy conservation are preserved.  These constraints are essential for
the stellar pulsation problem (\cf Buchler 1990; see also the tests below).
Indeed, an examination of the energetics shows the pulsation to be rather
delicate.  The pulsation kinetic energy ($\approx 8\times 10^{42}$ erg) is
generally very small compared to both the gravitational energy ($\approx
8\times 10^{47}$ erg) and the internal energy ($\approx 4\times 10^{47}$ erg)
which have opposite signs (from the virial theorem 2K + G $\approx$ 0).  In
addition, in the nonadiabatic case, the growth-rates are often small, \ie it
takes many pulsations to build up to the saturation amplitude.  In contrast,
the heat that flows through the stellar envelope in one pulsation ($\approx
2\times 10^{43}$ erg) is very large (The indicated numerical values are for a
typical classical Cepheid).  The coupling of the pulsation to the heat-flux
thus plays a very important role in the pulsations.  It determines both the
linear vibrational stability or instability and the saturation amplitude.
Needless to say, it therefore requires an accurate difference scheme.  One of
the important checks is that in the linear regime the growth-rates obtained
from the numerical hydrodynamics agree with those obtained from a linear
stability of the normal modes of oscillation.

The prescription for converting to an adaptive mesh is well known (\eg Winkler,
Norman \& Mihalas 1984).  Three velocities appear, namely $u$, the physical
velocity of the fluid with respect to a spatially fixed reference frame (here,
the equilibrium stellar model), $u_{grid}$, the velocity of the mesh points
with respect to the same reference frame and $u_{rel} = u - u_{grid}$ the fluid
velocity relative to the mesh.

The adaptive mesh hydrodynamic equations, in their usual notation,  are
given by

 \begin{equation}
{d\over dt} \int \rho d V + \int \rho u_{rel} dA  =0 
 \end{equation}
 \begin{equation}
{\delta (\rho\Delta V) \over \delta t} + \Delta (\rho u_{rel} A) = 0
 \end{equation} 

\vskip 10pt

 \begin{equation}
{d\over dt} \int u\rho d V + \int \rho u u_{rel} dA +
\int \left({\partial p \over \partial r} + \rho {GM\over r^2}
\right ) dV = 0
 \end{equation}
 \begin{equation}
{\delta (u\rho\Delta V) \over \delta t} + \Delta (\rho u u_{rel} A)
+ A\Delta p +{GM\over r^2} \rho\Delta V =0
 \end{equation}

\vskip 10pt

 \begin{eqnarray}
{d\over dt} \int \rho   \left( e + {u^2\over 2} -
  {GM\over r}\right ) dV
 &+& \int \rho \left (e + {u^2\over 2}
  - {GM\over r} \right) u_{rel} dA 
\nonumber \\ & & \\
 &+& \int \left ( pu + F_L\right ) dA = 0 \nonumber
 \end{eqnarray}

 \begin{eqnarray}
{1\over \delta t} \delta   \left [ \left (e + {u^2\over 2} -
  {GM\over r}\right ) \rho\Delta V\right ]
 &+& \Delta \left[ \left(
   e + {u^2\over 2} - {GM\over r}\right)\rho u_{rel} A\right]  
\nonumber \\ & & \\
 &+& \Delta\left[ \left (pu + F_L\right ) A\right ] = 0 \nonumber
 \end{eqnarray}

\ni where $\delta$ denotes the temporal change of a variable or expression and
$\Delta$ the spatial variation.  We treat the radiative flux in the diffusion
approximation as we have done in the past with the Lagrangean code.

These equations have to be completed with a mesh equation which specifies how
the mesh points are linked to the fluid.  There is of course no universal
prescription for a mesh equation, and we have adopted a variant of Dorfi \&
Drury (1987) which we shall describe in detail below.

In a separate paper we will relax the equilibrium diffusion approximation and
add separate radiation equations.

\section{\ \ The Code}

Fraley's scheme is a staggered Lagrangean scheme in which the basic radii,
velocities are defined at the cell edges ($i$), and the density and
thermodynamic variables are defined at the cell centers ($\ihp$) as shown
in Figure 1.

 \begin{figure}
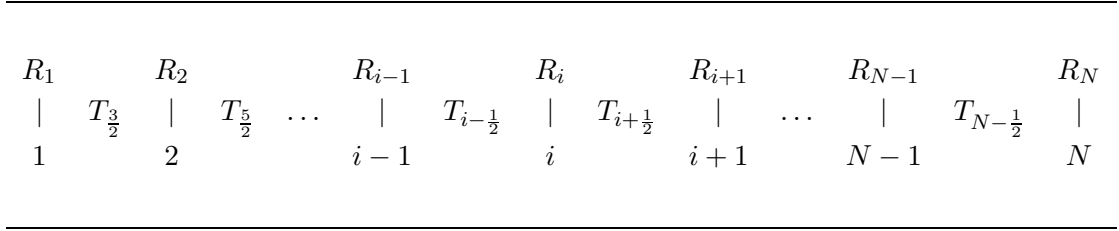

\vbox{
 \begin{center}
 \begin{tabular}{ccccccccccccccc}
\hline
&& \\
$R_1$ & & $R_2$ & & & $R_{i-1}$ & & 
$R_i$ & & $R_{i+1}$ & & $R_{N-1}$ & & $R_N$\\
$|$ & $T_{3\over2}$ &  $|$ & $T_{5\over2}$ & $\ldots$ &  $|$  & $T_\ihm$ &
$|$ & $T_\ihp$ &  $|$  & $\ldots$ & $|$ & $T_{N-{1\over2}}$ &  $|$  \\
 1  &   &   2  &    &       & $i-1$ &        & 
$i$ &        & $i+1$ &          & $N-1$ &     &  $N$\\
&& \\
\hline
 \end{tabular}
 \end{center}
}
\caption {The numerical  scheme}
 \end{figure}

Following common notation, each variable $y_i^n$ is assigned a spatial
subscript $i$ (or $\ihp$) and a temporal superscript $n$.  Intermediate
temporal values of the independent variables are defined with weights
$\theta_{_{y}}$.

 \begin{equation}
y^{n+\ahalf} = \theta_{_{y}} y^{n+1} + (1 - \theta_{_{y}}) y^n
 \end{equation}

The superscript (${\scriptstyle n}+\ahalf$ ) on a function of independent
variables, \eg the area $A^{n+\ahalf}$, thus denotes the temporal average of
the function.  Spatial differences are defined by $\Delta y_i = y_\ihp -
y_\ihm$ and $\Delta y_\ihp = y_{i+1} - y_i$, respectively.

Our 3$^{rd}$ equation is for the {\it total energy} (\ie the sum of the
internal, gravitational and kinetic energies) rather than for the {\it internal
energy} alone as in Fraley's scheme.  However, Fraley's scheme is such as to
actually conserve the global total energy exactly.  Therefore our differencing
reduces to Fraley's, to within numerical round-off errors when the grid motion
is Lagrangean ($u_{rel} = 0$).

\subsection{Continuity equation}

\ni With the usual definitions

 \begin{eqnarray} 
V_i^n  = {4\pi \over 3} (R^n_i)^3  \hbox{\hskip 1cm}
\Delta V^n_\ihp = V^n_{i+1} - V^n_i \nonumber
 \end{eqnarray}
 \begin{eqnarray}
A^n_i  = 4\pi (R^n_i)^2  \hbox{\hskip 1cm}
D\ngth M^n_\ihp  = \rho^n_\ihp \Delta V^n_{\; \ihp} \nonumber 
 \end{eqnarray}

\ni the continuity equation takes on the form

 \begin{equation}
D\ngth M^{n+1}_\ihp  - D\ngth M^n_\ihp 
    + F^\nhp_{M ,i+1} - F_{M ,i}^\nhp = 0,
 \end{equation}

\ni where $F_{M ,i}^\nhp$ is the mass flux through the zone boundary during the
timestep $\delta t$.  

There are many different ways of specifying the flux.  We have chosen the
second order scheme of van Leer (1979) for our code.  (We have not experimented
with higher schemes, \eg PPM; such schemes are much more cumbersome and
expensive to implement, but are not expected to increase the accuracy in the
shocks and ionization fronts which are already well resolved, although they
might increase the accuracy elsewhere).  In the following, the donor cell
scheme is used only for the purposes of comparison.

 \begin{equation}
\delta t \left (\rho u_{rel} A\right )^\nhp_i   =  F_{M, i}^{\;\; \nhp} 
   = \delta t A_i^\nhp \!\langle \rho_i^\nhp \!\rangle u_{rel,i}^{\;\; \nhp} 
 \end{equation}
 \begin{equation}
  \delta t u_{rel,i}^{\;\; \nhp} = \delta t u_i^\nhp - (R_i^{n+1} - R_i^n)
 \end{equation}
 \begin{equation}
\!\langle \rho_i^\nhp\!\rangle
   = \CS^-(u_{rel,i}) (\rho^\nhp_\ihp + d\rho^\nhp_\ihp) 
   + \CS^+(u_{rel,i}) (\rho^\nhp_\ihm - d\rho^\nhp_\ihm) 
 \end{equation}

\ni with the switch $\CS$ and $d\rho$ are defined as

 \begin{equation}
\CS^\pm (u)   = \ahalf (1 \pm sgn (u)) 
 \end{equation}
 \begin{equation}
d\rho_\ihp   = {\Delta\rho_{i+1} \Delta\rho_i \over
\rho_\ihpp - \rho_\ihm} \CS^+(\Delta\rho_{i+1} \Delta\rho_i).  
 \end{equation}

\ni For the purpose of comparison only we also use the less accurate first
order upwind scheme (donor cell):
 \begin{equation}
   \!\langle \rho_i^\nhp\!\rangle
  = \CS^-(u_{rel,i}) \rho^\nhp_\ihp + \CS^+(u_{rel,i}) \rho^\nhp_\ihm 
 \end{equation}

\subsection{Momentum Equation}

Again with the usual definition

 \begin{equation}
D\ngth M_i  = \ahalf \left (D\ngth M_\ihp + D\ngth M_\ihm\right )
 \end{equation}

\ni the two momentum source terms are differenced as

 \begin{equation}
{\cal M}_P  = \delta t \th (A\Delta p)_i  = 
\delta t \th\!\langle A_i\!\rangle ^\nhp
 \left ( p_\ihp^\nhp - p_\ihm^\nhp \right ) 
 \end{equation}
 \begin{equation}
{\cal M}_G  =\delta t\left( {GM\over r^2} \rho\Delta V\right)_i  
  = \delta t\th G\th\!\bigl\langle{1\over R_i^2}\!\bigr\rangle M_i^\nhp D\ngth
M_i^\nhp
 \end{equation}

In the Fraley scheme global energy conservation is possible through the
introduction of special averages.  We therefore introduce the same averages in
order to preserve this conservation in the Lagrangean limit

 \begin{equation}
\!\langle A_i \rangle^\nhp   = 
\fpi3 \Bigl( (R_i^n)^2 +R_i^{n+1}R_i^n +(R_i^{n+1} )^2\Bigl) 
 \end{equation}
 \begin{equation}
\!\Bigl\langle {1\over R_i^2}\!\Bigr\rangle   = {1\over R^{n+1}_i R_i^n} 
 \end{equation}

For the fluxes we again use van Leer's expressions

 \begin{equation}
\delta t (\rho u u_{rel} A)_i^\nhp 
    \equiv  F^\nhp_{I, \ihp}
= \delta t \langle u_\ihp^\nhp \!\rangle \delta M_\ihp 
 \end{equation}

\ni with 

 \begin{equation}
  \!\langle u_\ihp^\nhp\!\rangle 
   = \CS^-(\delta M_\ihp^\nhp) (u_{i+1}^\nhp + du_{i+1}^\nhp) 
   + \CS^+(\delta M_\ihp^\nhp) (u_i^\nhp     - du_i^\nhp)
 \end{equation}
 \begin{equation}
  du_i  = {\Delta u_\ihp \Delta u_\ihm \over
  u_{i+1} - u_{i-1} }\th\CS^+(\Delta u_\ihp \Delta u_\ihm)   
 \end{equation}
 \begin{equation}
  \delta M_\ihp  = \ahalf (\delta M_{i+1} + \delta M_{i}). 
 \end{equation}

\ni Finally the momentum equation becomes

 \begin{equation}
u_i^{n+1} D\ngth M_i^{n+1} - u_i^{n} D\ngth M_i^n  
   + \left(  F_{I,\ihp}^\nhp  
  -  F_{I,i -\ahalf}^\nhp \right) 
   + {\cal M}_P + {\cal M}_G = 0
 \end{equation}

There are different ways of introducing $\delta M_i$.  Algebraically it is
equal to $F^\nhp_{M,i}$, but numerically this form can cause problems (\cf
comments by Winkler \& Norman 1986, \S V.B.  on the use of $\delta M$).  The
direct replacement of $\delta M_i$ by the mass flux also widens the band matrix
in the iteration for each timestep as Eqs.~23 and 29 show.  Both problems are
avoided with the introduction of $M_i$ as an independent variable and its
definition

 \begin{equation}
\Delta M_\ihp = \rho_\ihp \Delta V_\ihp
 \end{equation}
 \begin{equation}
\delta M_i^\nhp = M_i^{n+1} - M_i^n
 \end{equation}

\subsection{{Total energy equation}}

A small complication arises because in Fraley's staggered scheme the internal
energies are most naturally defined in the cell centers, but the kinetic and
gravitational energies are defined at the cell boundaries.  We therefore define
the cell's total energy as being its internal energy plus one half of the
combined kinetic and gravitational energies of its edges.  
  \begin{equation}
 E_\ihp = e_\ihp D\ngth M_\ihp + g_{i+1} D\ngth M_{i+1} + g_i D\ngth M_i
  \end{equation}  \begin{equation}
 g_i  = \ahalf \Bigl( {(u_i)^2 \over 2}
    -{GM_i \over R_i  } \Bigr)
  \end{equation}

In accordance with Fraley's scheme we define the 'pressure-work'

  \begin{equation}
 (puA)_i^\nhp   =  \!\langle A_i^\nhp \!\rangle\th u_i^\nhp\;\; 
    \ahalf \left (p_\ihp^\nhp + p_\ihm^\nhp\right )
  \end{equation}

The total energy equation is

  \begin{equation}
 ( E_\ihp^{n+1} -  E_\ihp^n )
    + \delta t \th \Bigl( L_{i+1}- L_i
 +  (puA)_{i+1} -
     (puA)_i \Bigr )^\nhp 
  + \bigl(  F_{Ei+1}^{\;\;\, \nhp} -  
    F_{Ei}^{\;\;\, \nhp} \bigr)  = 0
  \end{equation}

\ni Note that the equation is written in full conservation form as a set of
spatial differences, $\Delta$.  We define the relative energy flux again in the
van Leer's fashion.  Care has to be taken to center the corresponding fluxes
correctly \ie consistently with Eq.~27. First we calculate the advection
terms of the kinetic energy at the cell centers, then calculate the spatial
average at the cell boundaries where the internal energy flux is naturally
defined.  Other kinds of averaging, such as first calculating the the total
energy density at cell centers and then advecting this quantity by $\delta M
_i^\nhp$, can lead to numerical instabilities.

  \begin{equation}
 F_{Ei}^\nhp   = \delta t \left [ 
 \!\langle e_i^\nhp \!\rangle \delta M _i^\nhp
 + \!\langle g_\ihm^\nhp \!\rangle \delta M _\ihm^\nhp
 + \!\langle g_\ihp^\nhp \!\rangle \delta M _\ihp^\nhp
 \right ] 
  \end{equation}
  \begin{equation}
  \!\langle e_i^\nhp\!\rangle
   = \CS^-(\delta M _i^\nhp) (e^\nhp_\ihp + de^\nhp_\ihp) 
   + \CS^+(\delta M _i^\nhp) (e^\nhm_\ihp - de^\nhp_\ihm) 
  \end{equation}
  \begin{equation}
  \!\langle g_\ihp^\nhp\!\rangle 
   = \CS^-(\delta M_\ihp^\nhp) (g_{i+1}^\nhp + dg_{i+1}^\nhp) 
   + \CS^+(\delta M_\ihp^\nhp) (g_i^\nhp     - dg_i^\nhp)
  \end{equation}
  \begin{equation}
  de_\ihp   = {\Delta e_{i+1} \Delta e_i \over
  e_\ihpp - e_\ihm} \CS^+(\Delta e_{i+1} \Delta e_i).  
  \end{equation}
  \begin{equation}
  du_i  = {\Delta g_\ihp \Delta g_\ihm \over
  g_{i+1} - g_{i-1} } \CS^+(\Delta g_\ihp \Delta g_\ihm)   
  \end{equation}

The luminosity $L$ is taken here in the radiation equilibrium diffusion
approximation

 \begin{equation}
L = - (4\pi r^2)^2 c {1\over\kappa (\rho ,T)} {d\over dm} {aT^4\over 3}
 \end{equation}

\ni We generally use the difference form suggested by Stellingwerf (1974) which
averages $T^4/\kappa$, but keep the option of using Stobie's (1969) average of
the opacity.

Finally a word about time-centering.  Fraley's scheme is {\sl stable} for
$\theta \ge \ahalf$.  It is therefore tempting to increase the stability of the
code by making all $\theta>0.5$.  However, in order to globally ensure total
energy conservation in the Lagrangean limit, it is necessary to use $\theta_u
=\ahalf$ for the velocity (Fraley 1968).  Furthermore, decentering the pressure
in Eqs.~(16) and (29) leads to improper mode growth or decay.  In fact, it can
be shown that even in the adiabatic limit the linear eigenvalues incorrectly
become complex (Kov\'acs, private communication); it is therefore important to
set $\theta_p =\ahalf$ as well.  Following Stellingwerf and our past experience
we use $\theta =\frac 23$ for the luminous flux $L$ which has a small effect on
the global results, except that is numerically stabilizing and reduces jitter.
(Aside from the special Fraley averages $\!\langle A \rangle$ and
$\!\langle1/R^2 \rangle$ the luminosity $L$ is therefore the only function that
is not centered in time).

In some calculations, for test purposes only, we replace the the total energy
difference equations (30) by the appropriate internal energy difference
equations, \viz

  \begin{eqnarray}
 ( U_\ihp^{n+1} -  U_\ihp^n )
    &+& \delta t \th \Bigl( ( L_{i+1}^\nhp - L^\nhp_i )  
 + p_\ihp \Delta \bigl(\langle A\rangle u\bigr )_\ihp \Bigr) \\
\nonumber \\ & & \\
  &+& \bigl(  F_{Ui+1}^{\;\;\, \nhp} -  
    F_{Ui}^{\;\;\, \nhp} \bigr)  = 0
\nonumber
  \end{eqnarray}

\ni where the internal energy is 
 $U_\ihp = e_\ihp \Delta M_\ihp$,
the internal energy flux
 $F_{Ui}^\nhp   = \delta t 
 \langle e_i^\nhp \!\rangle \delta M _i^\nhp$ with $\!\langle e_i^\nhp
\!\rangle$ 
 as defined in Eq.~32.

We note that this internal energy equation is also the equation solved by
Gehmeyr and by Dorfi \etal ({\it loc. cit.}).  We note in anticipation that the
performance of the code greatly deteriorates when this substitution is made.

\subsection{Pseudo-Viscosity}

Finally a word about pseudo-viscosity.  Our pressure everywhere includes a von
Neumann--Richtmyer (vNR) viscosity with a Stellingwerf cut-off parameter
$\alpha_q$, involving the local sound velocity $c$, of the form

 \begin{equation}
p_{visc,\ihp} = C_q \rho_\ihp \Bigl(\th\Bigl[ (u_{i+1}-u_i) 
     - \alpha_q c_{s,\ihp} \Bigr]_+\th\Bigr)^2
 \end{equation}

\ni In the case where spherical geometric effects are important this expression
is advantageously replaced by the form suggested by Whalen (\cf Buchler \&
Whalen 1990, Eqs. 9--10).

It may seem strange to still use pseudo-viscosity when the applied mathematical
literature abounds in 'better' ways of treating shocks (\eg Buchler 1990).
First, we note that the use of flux limiters and Riemann solvers (\eg Roe in
Buchler 1990) assumes that the hydrodynamics equations are written in a very
specific 'conservative' form which is not the case in our scheme, because of
our desire to preserve the good features of the Fraley scheme.  Eqs. 1--6 show
that in a Lagrangean or near Lagrangean situation all but the 'pressure-work'
flux and the radiative fluxes disappear and that the code would be unstable in
the absence of pseudo-viscosity.  Since, during the pulsation, parts of the
mesh can indeed behave in a manner close to Lagrangean the possibility of such
instability is not desirable.  In our 'defense' we also note that when the grid
motion redistributes mesh points into the shock, the effects of the
pseudo-viscosity are largely reduced.

\subsection{ The Mesh Equation}

Within constraints of stability, and ultimately accuracy, the choice of the
mesh motion is to a large extent an art.  On one extreme are the Eulerian and
Lagrangean descriptions with a 'rigid' mesh, with $u_{grid}=$constant and
$u_{rel}=0$, respectively, and on the other the mesh motions which follow the
dynamics so rapidly and faithfully that the physical quantities are largely
constant.  A good mesh-function clearly lies in between.

We have had good experience with Lagrangean meshes in which the outer zones,
say up to hydrogen ionization ($\approx$ 11,000\th K) were equally spaced and
the mesh size then increased geometrically inward.  In the spirit of our
conservative approach we therefore want to stay as close as possible to this
type of mesh both in the static model and in the hydro-code.  We want to use
the adaptive features primarily, first, to resolve the partial hydrogen
ionization during the pulsation, and second, to resolve shocks better than is
possible with a standard Lagrangean mesh.

For these reasons we adopt a mesh density defined in terms of the mass, rather
than the radius, \viz

 \begin{equation}
n_\ihp = {X_\ihp \over \Delta M_\ihp},
 \end{equation}

\ni but for tests we keep the option of using the radius in the mesh density
(\ie $n_\ihp = X_\ihp /\Delta R_\ihp$).  The quantity $X_\ihp$ is a mass scale
(or length scale).  In the case of periodic pulsators our methodology will be
to use first the fast Lagrangean Stellingwerf code to obtain the limit cycles
and then to switch over to an adaptive mesh with a minimum of transients.  For
this reason we use the cell masses $\Delta M_\ihp^{(0)}$ of the equilibrium
model for the scaling, \viz $X_\ihp = \Delta M_\ihp^{(0)}$.  This scale
$X_\ihp$ is thus held constant during the calculation.  With this definition of
$X_\ihp$ the grid density is uniformly equal to 1 in the equilibrium model and
in the Lagrangean code.  We note that in general the magnitude of $n_\ihp$
depends on the definition of $X_\ihp$ and really has a meaning only relative to
the values of the grid density at different zones; in other words, generally
the value of the grid density has nothing to do with the grid resolution.

Otherwise we adopt the mesh diffusion (parameter $\alpha$) and relaxation
time ($\tau$) of Dorfi \& Drury (1987):

 \begin{equation}
 \hat n_\ihp = n_\ihp -\alpha (1-\alpha) (n_{\ihpp} -2n_\ihp +n_\ihm )
 \end{equation}

 \begin{equation}
 \tilde n_\ihp^{n+1} = \hat n_\ihp^{n+1}
   + {\tau\over\delta t} (\hat n_\ihp^{n+1}-\hat n_\ihp ^n)
 \end{equation}

\ni For the structure function which gives the necessary grid resolution we
have found it convenient to use the following function:

 \begin{equation}
 Z_\ihp = \left(1 
  + \sum_k \beta_k S_\ihp 
  \left({\Delta f^{(k)}_\ihp\over F^{(k)}_\ihp}\right)^2
   \right)^{1/2},
 \end{equation}
where the scale factor
 \begin{equation}
 S_\ihp = {n_\ihp \over 1 + (s\th\th n_\ihp)^4}
 \end{equation}

\ni is introduced to give an upper limit to the grid density, \viz $1/s$.
Without the fourth order term in the denominator the grid density could
increase unlimitedly causing numerical problems at velocity discontinuities:
The artificial viscosity (Eq.~38) puts a constant number of grid points in the
shock region, a number that is governed by the parameter $C_q$.  This leads to
constant velocity difference $\Delta u$ independently of the current grid
density.  If then the $\Delta f^{(k)}_\ihp / F^{(k)}_\ihp$ term in Eq.~42 is
dominated by $\Delta u$, the larger value of grid density generates a larger
value of $Z_\ihp$ and causes another increase of $n$.  This positive feedback
is stopped by the upper limit of grid density.

Generally we introduce two terms in the sum (42), \viz

 \begin{equation}
 \Delta f^{(1)}_\ihp   = T_\ihpp-T_\ihm, \quad\quad F^{(1)}_\ihp = T_\ihp
 \end{equation}
 \begin{equation}
 \Delta f^{(2)}_\ihp   = u_{i+1}-u_i, \quad\quad F^{(2)}_\ihp = u_o
 \end{equation}

\ni Here the quantity $u_o$ represents the velocity scale of the problem which
can be mesh and time dependent.  The $f^{(1)}$ function serves the purpose of
placing meshpoints in the regions of rapidly varying temperature, such as in
the partial hydrogen ionization region which has a very sharp temperature
variation, particularly in classical Cepheids.  The $f^{(2)}$ function
acts only when strong velocity gradients are present and its obvious
purpose is to resolve shocks.

\ni With these preambles the mesh equation finally is

 \begin{equation}
 {\tilde n_\ihp \over Z_\ihp} = {\tilde n_\ihm \over Z_\ihm}
 \end{equation}

The specification of our mesh function is incomplete without boundary
conditions.  First we impose that the mesh be Lagrangean in the first few inner
zones and in the last few zones.  Second, following Dorfi \& Drury (1987) we
impose that $n_\ihm=n_\ihp$ at both Lagrange-adaptive interfaces.

With our definition of the grid density ($n_\ihp = \Delta M_\ihp^{(0)} / \Delta
M_\ihp$) the grid motion is Lagrangean when $\beta_k=0$ because $Z_\ihp=1$ and
$\Delta M_\ihp$ remains constant.  In other words the mesh equation is
always satisfied when the $\beta_k$ parameters are set to zero and the mass
distribution is given by the initial Lagrangean model ($Z_\ihp = n_\ihp = 1$).
By a gradual increase of the $\beta_k$ it is possible to evolve the Lagrange
grid into the adaptive one (\cf also \S 3.3).

For test purposes we have also experimented with the radius differences in the
grid density with $X_\ihp = \Delta R_\ihp^{in}$ (\ie $n_\ihp=\Delta R_\ihp^{in}
/ \Delta R_\ihp$).  In that case the $\beta_k=0$ condition gives a simplified
grid equation: $\Delta R_\ihp / \Delta R_\ihm =$constant, which defines a {\sl
quasi Eulerian} grid motion.  Since several zones at the top and the bottom of
the model are Lagrangean, the grid motion of these special zones are given by
the fluid motion: $R_{gr,L}^n = R_L^n$. With these boundary conditions the
solution of the simplified grid equation becomes
 \begin{equation}
 R_i^n = R_{L_1}^n + {R_i^0 - R_{L_2}^0 \over R_{L_2}^0 -R_{L_1}^0}
                  (R_{L_2}^n - R_{L_1}^n),
 \end{equation}
 where $R_{L_1}$ and $R_{L_2}$ are the radius variations of the two Lagrangean
zones bracketing the quasi Eulerian grid.

\section{Tests}

As performance tests for the code we have chosen the challenging Noh spherical
shock problem (Noh 1987) and Sedov's (1959) point explosion problem.
Analytical solutions exist in both cases.  A perfect gas equation of state with
($\gamma=5/3$) has been used for these tests.  We will also present
calculations for Cepheid model pulsations with a realistic equation of state
and opacities to show the application of the code to stellar oscillations.

\subsection{Spherical Noh Problem}

The initial conditions for the Noh problem are uniform velocity ($u_j$=--1) and
density ($\rho_j$=1) with zero pressure and temperature.  The shock generated
at the center of the sphere moves outward with a constant speed ($v$=1/3).  The
density has a flat profile behind the shock with $\rho$=64 at all times.

 \begin{figure}
 \hskip-0.5truecm\psfig{figure=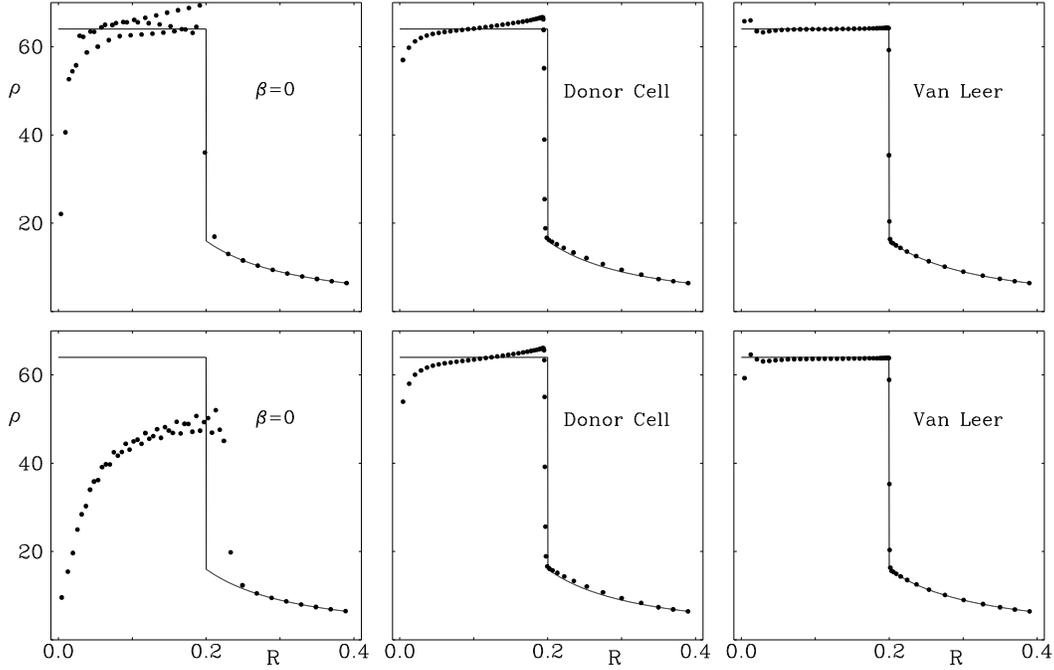,width=14.truecm}
 \caption {Numerical solution of the spherical Noh problem at t=0.6. 
 Upper panels: tensor viscosity, lower panels: vNR viscosity, 
 $\Delta M$ is used in the mesh density.}
 \end{figure}

 \begin{figure}
 \hskip-0.5truecm\psfig{figure=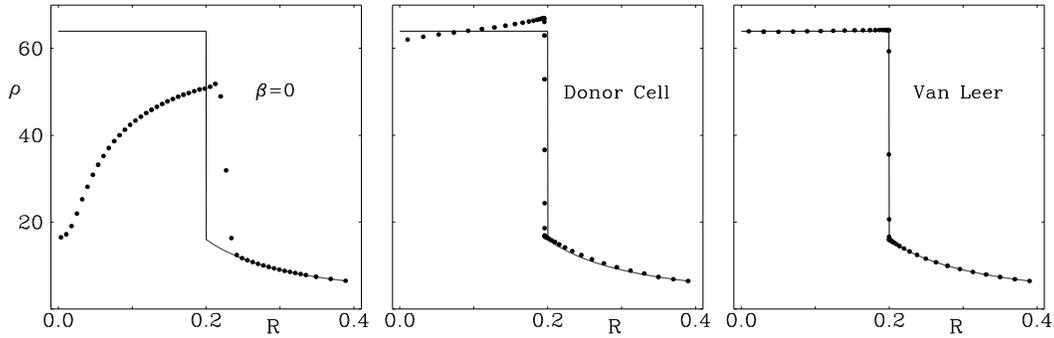,width=14.truecm}
 \caption {Spherical Noh problem.  $\Delta R$ is used in the grid density; vNR
viscosity.}
 \end{figure}

 \begin{figure}
 \hskip-0.5truecm\psfig{figure=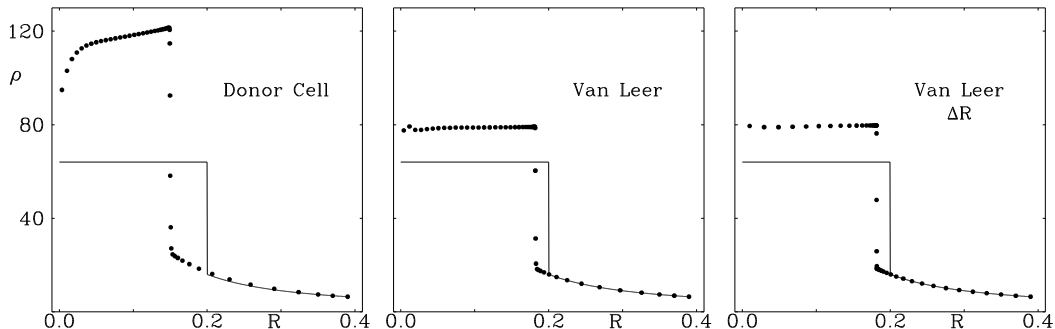,width=14.truecm}
 \caption {Spherical Noh problem.  The internal energy eq.  is solved instead
of the total energy eq.; vNR viscosity}
 \end{figure}

For the numerical solution we have used a mesh with 50 zones initially spaced
with equal $\Delta R$.  Since the spherical geometric effects are important we
have also used the tensor artificial viscosity for these calculations (For a
comparison of different treatments of artificial viscosity in Lagrangean
calculations see Buchler and Whalen 1990).  The parameters of the calculations
are the following: $C_q$=1.0, $\beta_u$ = 1000, $s$=0.0002, $\tau$=0.0005,
$\alpha$=2.0 and $u_o=1$.

Figure 2. displays the results of the calculations for different parameters
together with the analytic solution.  For the upper and lower row the Whalen
tensor and the von Neumann-Richtmyer viscosity have been used, respectively.
The adaptive calculation with the first order upwind scheme (second column)
considerably improves the solution over the Lagrangean limit ($\beta_u$=0)
which exhibits unphysical oscillations behind the shock.  The use of the van
Leer advection provides superior results (third column).  This test clearly
demonstrates that the higher order advection scheme is preferable in the
adaptive calculation.  The results are independent of the choice of the
artificial viscosity, in fact the use of the tensor artificial viscosity is not
necessary with the adaptive computation even in this problem with strong
sphericity.

In Fig. 3. we present the solutions with an alternate form for the grid
density, \viz with $\Delta R_\ihp$ in the denominator in Eq.~39.  The vNR
artificial viscosity is used.  The first panel shows the quasi Eulerian result
($\beta_u$=0).  The result with $\Delta R_\ihp$ is a little better than the one
with $\Delta M_\ihp$, especially at the center, but the difference is
insignificant elsewhere.  (The glitch near the center is due to the
inaccuracies introduced by the steeply varying cell mass in the initial profile
which has been constructed with equal $\Delta R$.)

We have to emphasize that solving the {\sl total energy equation} (Eq.~30) has
been crucial in getting a correct solution for the Noh problem.  We illustrate
this in Figure~4 which shows the results for a calculation in which we now
solve the internal energy equation (Eq.~437).  It is true that with the adaptive
mesh code we still get a sharp shock, but the velocity of the shock front (and
thus the density) is very different from the correct values.  When we monitor
the time-dependence of the energies we notice a small gradual decrease in the
global total energy (4.2\% at t=0.6).  Furthermore, the kinetic energy is too
large and correspondingly the internal too small by $\approx 7.3\%$ and 15.6\%
respectively, whereas with the solution of the total energy equation the
discrepancy in kinetic energy is less than 0.5\%, \ie 10 times less.  In the
Lagrangean limit, of course, the two energy equations yield the same solution
to machine accuracy (because of the already mentioned property of the Fraley
scheme).

The order of the advection scheme plays a much more important role in the
accuracy of the calculations when the internal energy equation is solved, as a
comparison of Fig.~2 (top middle and top right) and Fig.~4 (left and middle)
shows.  This is an indication that the advection errors are much larger for the
internal energy equation.  An additional reason for the inaccuracies is found
in the differencing of the source term (Eq.~37)
 $p \Delta \bigl(\langle A\rangle u\bigr )$
 which can vary very rapidly throughout the shock.  In contrast, the total
energy equation contains only flux terms and the corresponding
Hugoniot--Rankine equation is automatically satisfied through
near-discontinuities.

The middle and right graphs of Figure~4 show that the inaccuracies associated
with the solution of the internal energy equation occur for both definitions of
mesh density (mass \vs radius).

 \subsection{Sedov Problem}

 \begin{figure}
 \hskip-0.5truecm\psfig{figure=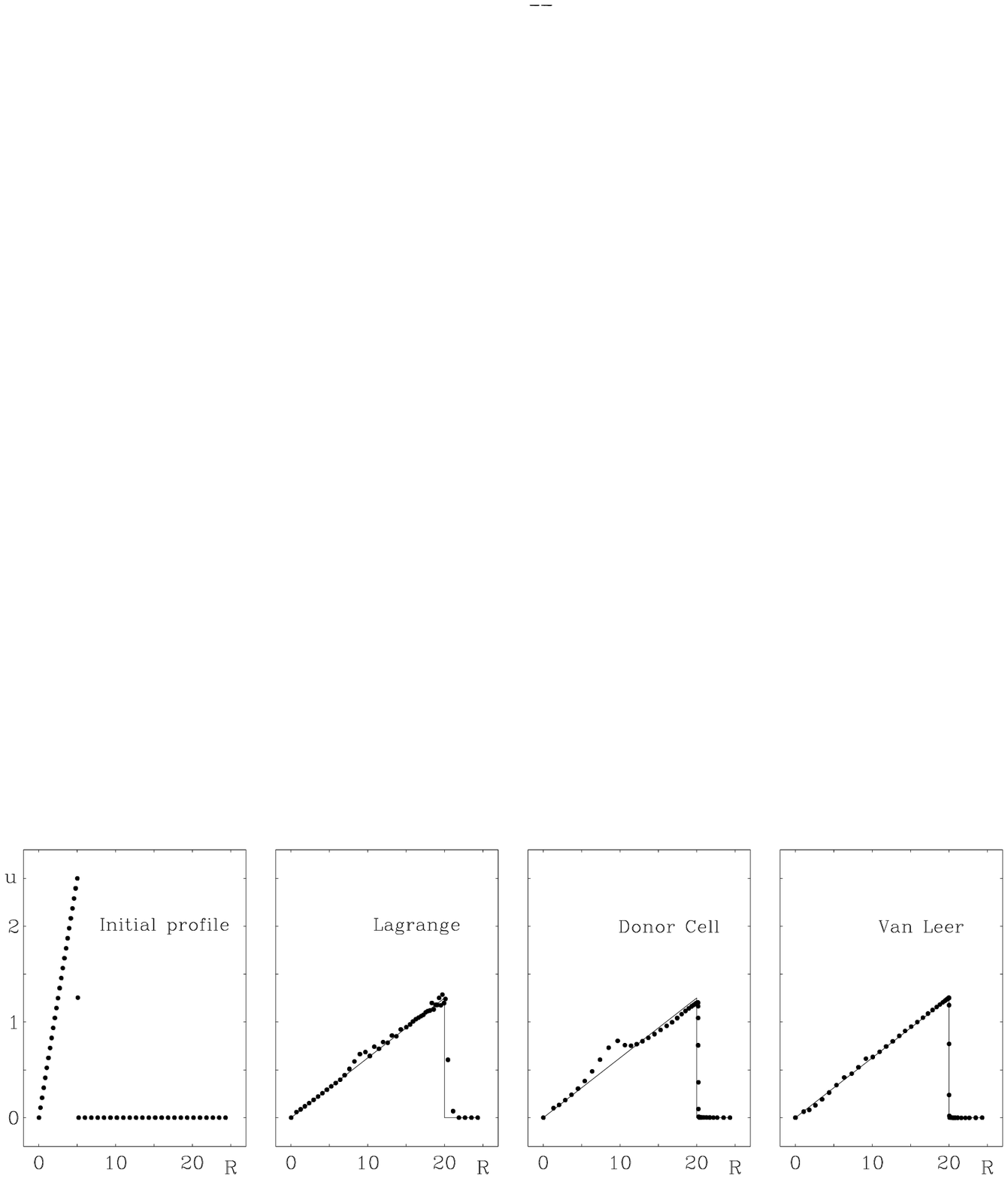,width=14.truecm}
 \caption {Numerical solution of the Sedov problem. 
 From left to right: initial velocity profile, Lagrangean and
 adaptive results.
 }
 \end{figure}

The analytic solution of an intense point explosion in a power-law density
gradient with spherical symmetry is given by Sedov (1959).  The motion of the
gas is self-similar \ie the shapes of the pressure, density and velocity
profiles remain the same during the explosion.  For an ideal gas with
$\gamma=5/3$ the distance of the shock front from the center ($R_s$) is
proportional to $t^{2/3}$, and the physical parameters behind the shock are
given by the initial value of the pressure, density and shock velocity.  For an
initial density given by $\rho(R) = A R^{-2}$ the solution has the form

 \begin{eqnarray} u = {R \over 2 t} \hbox{\hskip 1cm} \rho = {4
 A R \over R_s^3 } \hbox{\hskip 1cm} p = {A \over 3 t^2} \left ( {R
\over R_s}
 \right )^3,
 \end{eqnarray} 

\ni for radii $R$ less than the radial distance of the shock ($R_s$).  The
energy of the explosion appears in these functions only through the location of
the shock front ($R_s$) for a given time.

We choose the explosion energy so that the position of the shock at $t$=1 is
$R_s$=5 (in arbitrary units).  Our initial model is constructed by discretizing
the exact solution for $t$=1.  Half of the 50 zones are placed below the front.
The zones are initially equally spaced in $\Delta R$, but with different
spacings below and above the front to get approximately the same zone mass on
both sides.  We use $\Delta M$ in the grid density.  The numerical parameters
are $C_q$=1.0, $\beta_u$ = 1000, $s$=0.005, $\tau$=0.003, $\alpha$=2.0 and
$u_o=1$.

The Lagrangean and adaptive solutions at $t$=8 are shown in Figure 5 for the
van Leer scheme.  For comparison only Donor Cell results are also shown.  By
this time the intense shock, with density ratio 4:1 across the front, has
progressed to a density 1/16 times the density at the start of the calculation.

When the initial grid distribution does not satisfy the mesh equation, which is
the general situation, transient oscillations appear in the solution.  Figure~5
shows that the initial transient leaves a large incorrect permanent scar on the
solution when the inaccurate donor cell advection scheme is used.  Only
when we use the second order scheme are the fluctuations in the velocity
minimal, and the position of the shock front is correctly obtained.

As for the Noh problem, we again find it important to solve the total energy
equation rather than the internal energy equation.  With the use of the
internal energy equation, the resulting shock is still sharp, but it lags.  For
example, at $t$=8 the position of the front is $R_s$=19.4 instead of the
correct value of $R_s$=20.  The accumulated error in the total energy reaches
10\% for this time.  The total kinetic energy which should be constant during
the shock propagation is also decreasing at the same rate.  It is this
underestimation of the kinetic energy that causes the error in the shock
position.

 \subsection{Cepheid Pulsations}

 \begin{table}[h]
 \caption[]{Typical errors of the total energy -- Cepheid model problem.}
 \begin{center}
 \begin{tabular}{ccccc}\hline
 $j_L$  &$T(j_L)$ [K] & $\epsilon_A(E_{tot})$ 
   & $\epsilon(E_{tot})/P$ & Scheme \\
 \hline
 60 & 5 10$^4$ & 4.0 $10^{-9}$ & --1.0 $10^{-9}$ & van Leer \\
 40 & 1 10$^5$ & 1.5 $10^{-7}$ & --1.5 $10^{-8}$ & van Leer \\
 30 & 2 10$^5$ & 4.0 $10^{-7}$ & +1.0 $10^{-8}$  & van Leer \\
 60 & 5 10$^4$ & 3.0 $10^{-8}$ & --6.0 $10^{-8}$ & Donor Cell \\
 40 & 1 10$^5$ & 4.0 $10^{-6}$ & --5.0 $10^{-6}$ & Donor Cell \\
 \hline
 \end{tabular}
 \end{center}
 \end{table}

 \begin{figure}
 \hskip-0.5truecm\psfig{figure=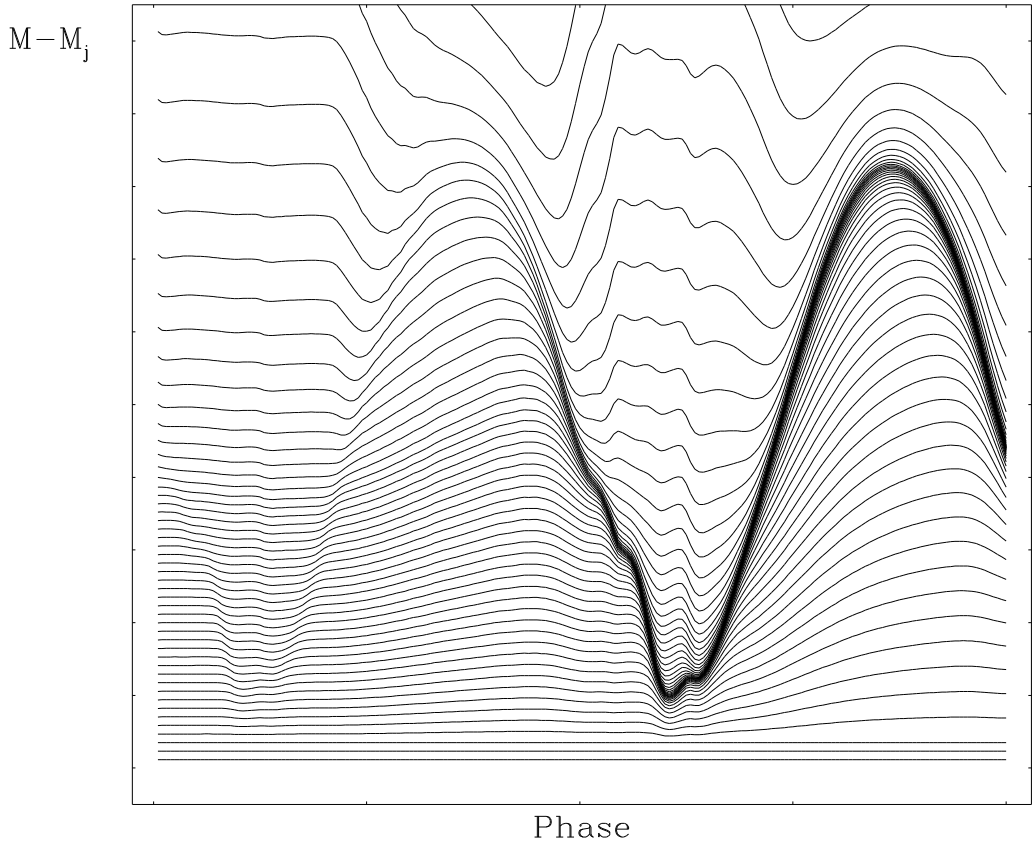,width=14.truecm}
 \caption {Transition from Lagrange to adaptive grid.}
 \end{figure}

We have selected a model from one of the Cepheid sequences of Moskalik, Buchler
and Marom (1992).  We use the OPAL opacities (Iglesias \etal 1992) matched with
the opacities of Alexander and Ferguson (1994) at low temperatures.  The
chemical composition is $X$=0.7, $Z$=0.02.  The computations are performed with
120 mass zones.  The static model has been built as follows: equal mass zones
inward from the surface; a temperature anchor of 11,000K in the 30th inward
zone; the mass of the inner zones increases geometrically up to an inner
temperature of $T_{in}=2\times 10^6$ K (This is the same type of grid that has
been found to give good results in the Lagrangean code, \eg Moskalik \etal
1992).  This mass distribution also gives the reference grid for the adaptive
calculations by setting the values of $X_\ihp$, as discussed earlier.  With the
stellar parameters $M$=6.5\Mo, $L$=7213\Lo\ and \Teff=5404 K the period of the
model is $\approx$ 18 days.

 \begin{figure}
\hskip-0.5truecm\psfig{figure=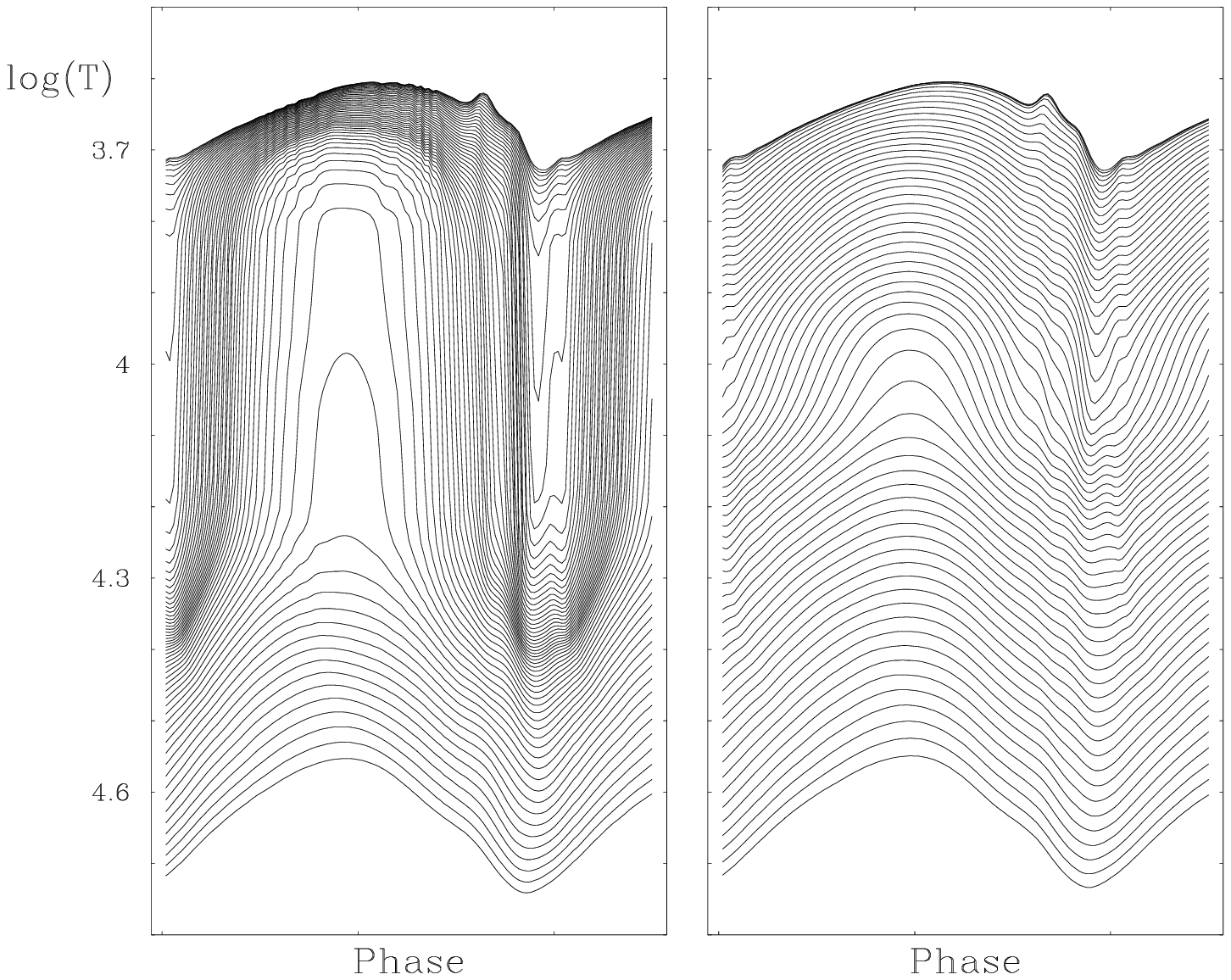,width=14.truecm}
\caption {Temperature profiles for the bump-Cepheid model Left box:
Lagrangean grid, right box: adaptive grid}
 \end{figure}

It is notoriously difficult to get rid of long lasting transient oscillations
in an adaptive code.  We have found the following procedure to be very useful
in the case of periodic pulsations (asymptotic limit cycle).  A limit cycle
solution is first calculated with the relaxation method of
\hyphenation{Stellingwerf} Stellingwerf applied to the Lagrangean code
(starting from a perturbed static equilibrium model).  This pulsating model is
then inserted into the adaptive code in its Lagrangean mode, and the adaptive
mesh is gradually switched on.  More specifically, the $\beta_k$ parameters of
the structure function are increased during the first $s$ periods according to
the following recipe:
 \begin{equation}
 \beta_k = 10^{3t/sP - 3} \beta_k^o,
 \end{equation}
 and they are then held constant for $t > sP$.  This smooth switching on of the
grid parameters transforms the Lagrangean mesh into an adaptive one.  Since the
grid equation is satisfied at the beginning of the calculation (\ie in the
$\beta_k$=0, Lagrangean limit) {\sl we can evolve the model from Lagrangean to
adaptive smoothly by this method} and avoid annoying and often very slowly
decaying transients.  When the advection errors are small (see below) the code
converges from a Lagrangean to an adaptive grid limit cycle in several tens of
periods.

The transition of the grid from Lagrangean to adaptive is shown in Figure~6 for
$s=2$.  The positions of the zones are given by $M -M_j$ ($j$=70--120).  The
location of the partial ionization region is clearly revealed by the higher
grid density which it attracts during the second period of pulsation.

The temperature parameter of the structure function is set to $\beta_T^o =
$100.  For most of the tests we have used $\beta_u^o =$0 (the resolution in
temperature is enough to get a smooth variation), but we also have made
calculations with $\beta_u^o =$10 and $u_o=10^5$cm/s.  This results in a
sharper shocks, but too large a value of $\beta_u^o$ attracts too many points
into the shock and away from the partial ionization front, so that the latter
loses its sharpness.  The given values represent a good compromise between the
two tendencies for this number (120) of total mesh points.  For the other
parameters we have chosen the values $s$=0.005, $\tau$=0.1 days and
$\alpha$=2.0.  The pseudo viscosity in the form of Eq.~43 is used with $C_q$=4
(for some tests $C_q$=1) and $\alpha_q$=0.01.

 \begin{table}[h]
\caption[]{Fourier parameters of the limit cycle solution -- radius 
variation. (LA: Lagrangean; TE: total energy, IE internal energy Eqs. are 
 solved; vL: van Leer, DC: donor cell.}
 \begin{center}
 \begin{tabular}{cccccccccc}\hline
$j_L$ & $C_q$ &  & Scheme & 
$P_0$ & $A_1$ & $R_{21}$ & $R_{31}$ & $\phi_{21}$ & $\phi_{31}$ \\
\hline
-- & 4 & LA & -- & 17.6 & 22.6 & 0.193 & 0.017 & 6.159 & 0.725 \\
-- & 1 & LA & -- & 17.6 & 23.2 & 0.197 & 0.021 & 6.176 & 0.906 \\
\hline
60 & 4 & TE & vL & 17.6 & 22.5 & 0.197 & 0.026 & 6.190 & 0.534 \\
60 & 4 & IE & vL & 17.6 & 22.5 & 0.195 & 0.025 & 6.194 & 0.569 \\ 
60 & 1 & TE & vL & 17.6 & 23.1 & 0.199 & 0.027 & 6.200 & 0.824 \\ 
60 & 1 & IE & vL & 17.6 & 23.1 & 0.199 & 0.027 & 6.203 & 0.842 \\ 
\hline
40 & 4 & TE & vL & 17.5 & 22.1 & 0.186 & 0.024 & 6.213 & 0.590 \\
40 & 4 & IE & vL & 17.5 & 21.3 & 0.182 & 0.017 & 6.177 & 0.307 \\
40 & 4 & TE & DC & 18.7 & 35.4 & 0.347 & 0.159 & 6.265 & 6.121 \\ 
40 & 4 & IE & DC & 18.0 & 17.0 & 0.148 & 0.070 & 0.395 & 1.176 \\ 
\hline
30 & 4 & TE & vL & 17.3 & 17.5 & 0.138 & 0.033 & 0.127 & 1.207 \\ 
30 & 4 & IE & vL & 17.3 & 16.6 & 0.130 & 0.018 & 0.021 & 1.185 \\ 
\hline
 \end{tabular}
 \end{center}
 \end{table}

 \begin{table}[h]
 \caption[]{Fourier parameters of the limit cycle solution -- magnitude
 variation. (LA: Lagrangean; TE: total energy, IE internal energy Eqs. are 
 solved; vL: van Leer, DC: donor cell.}
 \begin{center}
 \begin{tabular}{cccccccccc}\hline
 $j_L$ & $C_q$ &  & Scheme & 
 $P_0$ & $A_1$ & $R_{21}$ & $R_{31}$ & $\phi_{21}$ & $\phi_{31}$ \\
 \hline
 -- & 4 & LA & -- & 17.6 & 0.41 & 0.307 & 0.250 & 5.815 & 5.324 \\
 -- & 1 & LA & -- & 17.6 & 0.42 & 0.310 & 0.250 & 5.837 & 5.328 \\
 \hline
 60 & 4 & TE & vL & 17.6 & 0.41 & 0.299 & 0.229 & 5.817 & 5.202 \\
 60 & 4 & IE & vL & 17.6 & 0.41 & 0.300 & 0.229 & 5.825 & 5.212 \\
 60 & 1 & TE & vL & 17.6 & 0.42 & 0.301 & 0.248 & 5.784 & 5.226 \\
 60 & 1 & IE & vL & 17.6 & 0.42 & 0.301 & 0.249 & 5.790 & 5.227 \\
 \hline
 40 & 4 & TE & vL & 17.5 & 0.40 & 0.276 & 0.211 & 5.878 & 5.089 \\
 40 & 4 & IE & vL & 17.5 & 0.38 & 0.282 & 0.209 & 5.851 & 5.162 \\
 40 & 4 & TE & DC & 18.7 & 0.64 & 0.364 & 0.144 & 0.049 & 6.082 \\
 40 & 4 & IE & DC & 18.0 & 0.31 & 0.209 & 0.171 & 5.712 & 4.461 \\
 \hline
 30 & 4 & TE & vL & 17.3 & 0.32 & 0.241 & 0.151 & 5.792 & 4.861 \\
 30 & 4 & IE & vL & 17.3 & 0.30 & 0.248 & 0.143 & 5.778 & 4.990 \\
 \end{tabular}
 \end{center}  \end{table}

 \begin{figure}
 \hskip-0.5truecm\psfig{figure=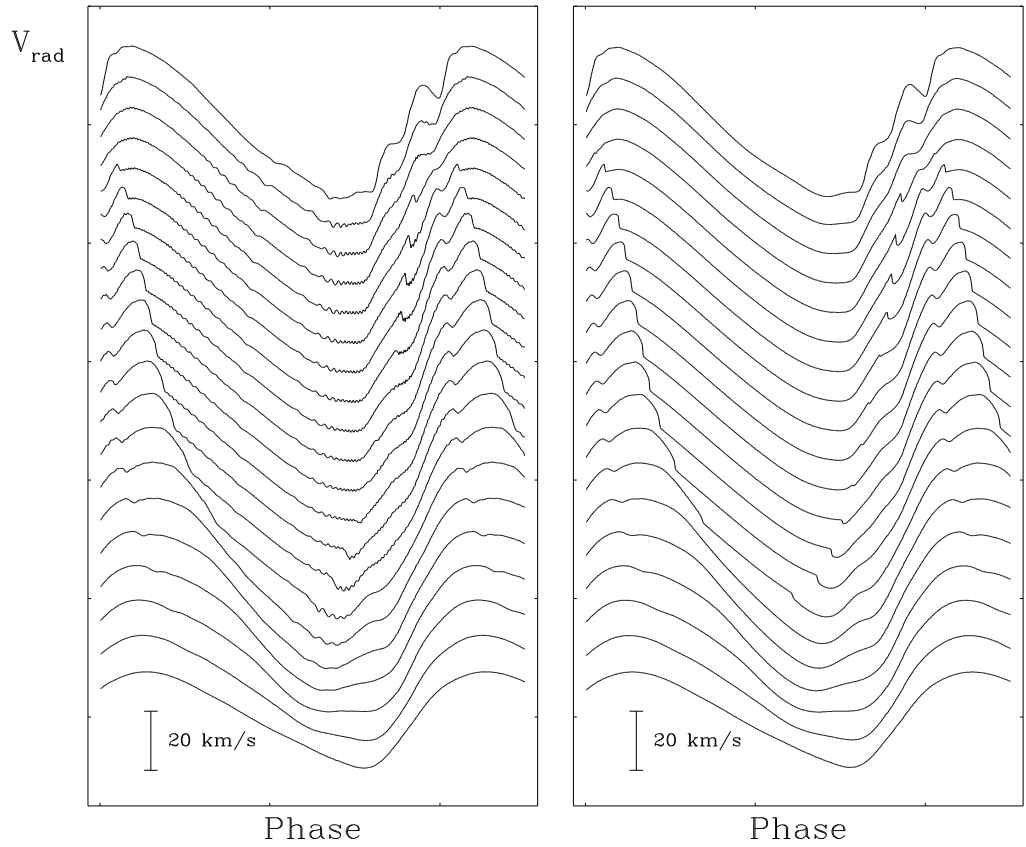,width=14.truecm}
\caption {Velocity profiles for the bump-Cepheid model. Left:
Langrangean, right: adaptive grid.}
 \end{figure}

 \begin{figure} 
 \hskip-0.5truecm\psfig{figure=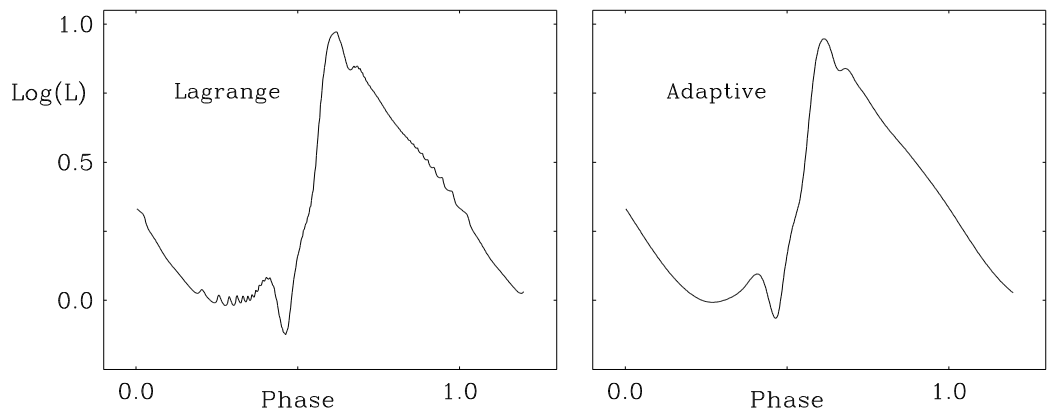,width=14.truecm}
 \caption {Light-curve. Left: Langrangean, right: adaptive grid.}
 \end{figure}

The spatial switch-over from Lagrangean to adaptive mesh occurs at zone $j_L$ .
When it is made too far inside the stellar model (small $j_L$) we experience a
numerical instability.  This happens whether we solve the total energy equation
or the internal energy equation.  We note that the same problem has been
encountered by Feuchtinger \& Dorfi (1994).  The reason ultimately lies in the
mesh equation.  The global nature of the mesh equation and the mesh motions
that occur in the outer zones because of moving physical features (mostly the H
ionization front) induce small mesh motions even in the relatively quiescent
inner zones.  However, because of the large masses of these zones the internal
and total energy advection errors can therefore reach the order of the small
pulsation energy itself.  The indication that the trouble is caused by
advection errors can be glimpsed from Table 1.  Here we report results obtained
with the use of the internal energy equation (Eq.~37) because this lets us then
monitor total energy conservation.  We show both the long-term global error
$\epsilon (E_{tot})$, defined as the sum of the errors per cycle of all the
zones divided by ${E_tot}$, and the amplitude of the error fluctuations during
the cycle $\epsilon_A(E_{tot})$ (also normalized by $E_{tot}$).  (All the
errors are below the convergence parameter, $10^{-5}$, of the internal
iterations of the code.)  The long term error is indeed seen to be
substantially larger with the first order advection scheme.  When the first
order scheme is used, the variation of the energy error is dominated by the
long term trend.

The solution to the numerical instability is to make the motion of the inner
envelope Lagrangean, as Feuchtinger \& Dorfi (1994) have also found.  The
selection of the location of the switching point $j_L$ between adaptive and
Lagrangean mesh is important, though, as Table~1 shows.  An optimal switch
$j_L$ should be as far out in the envelope as possible, but above the He II
ionization (at $T>5-10\times 10^4$ K).

Unfortunately no suitable analytical results exist for a stellar pulsator model
against which one could check the codes.  We therefore have to rely on the
robustness of the results with respect to numerical parameters on the one hand,
and to the agreement that we can achieve with the observational data.  Although
it is important that the pulsations should exhibit long term stability (total
energy and pulsation energies constant), this, by itself, is not a guarantee
that the amplitude of the cycle is in fact correct.

In order to compare the robustness of the computed limit cycle pulsation we
display their Fourier parameters, for the radial velocity of the photo-sphere
in Table II and for the stellar magnitude in Table III.

First we note that the calculations with the new Lagrangean code confirm the
results of Moskalik, Buchler and Marom (1992) which, we recall, gave Fourier
parameters of the bump Cepheid models that were rather close to the observed
values.

Second, it immediately stands out that it is imperative to use a second order
scheme for the fluxes, confirming the results of our test examples.  We note
parenthetically that in a sequence of Cepheid models that we calculated the
donor cell fluxes produced results with trends in the amplitude \vs period
behavior that were completely wrong, even opposite to those both of the higher
order scheme and of the purely Lagrangean scheme.  For the models calculated
with Donor cell advection scheme, the shifts of amplitudes are even opposite
for the two choices of handling the energy.

Third, the Fourier parameters exhibit a reasonable robustness when the spatial
switchover from Lagrangean to adaptive mesh occurs in the range $j_L$=40--60.
Values of $j_L$=30 give already unsatisfactory results.  Note also that when
the internal energy (IE) equation is solved the Fourier parameters are more
sensitive to $j_L$ because of the increased advection errors.  This is
especially true for the third order radial velocity coefficients $R_{31}$ and
$\phi_{31}$.  This again confirms the numerical superiority of the total energy
over the internal energy equation.

Fourth, the adaptive code although requiring some pseudo-viscosity, lets one
reduce pseudo-viscous dissipation below values that are possible with the
Lagrangean code.  The increase of the pulsation amplitude with $C_q$ clearly
shows that this amplitude is largely determined by the artificial viscosity
independently of the choice of grid.  This again illustrates the well known
problem that when we reduce the numerical dissipation below the 'standard'
values, the pulsation amplitudes get larger than the observed ones (for a
discussion \cf \eg Kov\'acs 1990).  It must be considered a lucky coincidence
that, since the pioneering work of Christy in the 1960s, the use of the
'standard' vNR pseudo-viscosity has yielded pulsation amplitudes in rough
agreement with the observed ones.  In order to make progress it will be
necessary to replace the pseudo-viscous dissipation with the correct physical
one, most likely eddy viscosity.  It is very likely that the violent shocks
induce turbulent motions, especially in the relatively 'soft' (small $\Gamma$)
partially ionized regions.  But to even get a good understanding of where in
the star and when in the pulsation phase such turbulent dissipation plays a
role, and to get even a crude idea of itsw magnitude, one needs a 3D
hydrodynamic simulation.

Finally, we note that the great advantages of the adaptive numerical
calculations is the smoothness of the results.  In Figure 7. the temperature
variations of the outer 60 shells are shown both for the Lagrangean and for the
adaptive results (in both cases as a function of the respective grid indices).
The resolution of the spatial temperature structure is clearly superior by the
adaptive mesh.

In Fig. 8 we present the fluid velocity variations of different layers of the
model (spread out vertically for clarity), both for the Lagrangean and for the
adaptive results.  The plotted masses have been selected at the values
$M^{L}_{60}$, $M^{L}_{63}$, $M^{L}_{66}\dots$ $M^{L}_{120}$ of the Langrangean
model.  In order to compare we have had to interpolate the velocities from the
adaptive code for these given values of the masses.  The adaptive code is seen
to provide significantly smoother curves than the Lagrangean  The relatively
sharp features on some of the velocity curves indicate the position of the
hydrogen partial ionization front.

Finally, Figure~9 shows the improvement in the quality of the 
light-curve that
is obtained with the adaptive mesh.

\section{Conclusion}

We have described an adaptive hydrocode which has been specifically developed
for stellar pulsation calculations.  The code has been tested for shock
problems with known analytical solutions.  Purposedly, to mimic more realistic
calculations in which few mesh points are available for the shock by itself, a
relatively coarse mesh has been used.  These tests lead to the following
conclusions:

\ni -- The first order schemes (donor cell) do not give satisfactory
agreement between the numerical and analytical solution.  At least a second
order (\eg van Leer) scheme is necessary.

\ni -- The solution of the total energy equation is preferable to the use of an
internal energy equation because the numerical deterioration of the results due
to the advection errors is reduced.

\ni -- A careful application of the adaptive grid technique can provide
superior results for tough shock problems even with a relatively few grid
points.

It is of course much easier and much faster to run a Lagrangean code than an
adaptive one.  In particular, the Stellingwerf Lagrangean relaxation method is
very efficient when periodic cycles are to be computed as for classical
Cepheids or RR Lyrae.  The fact that the adaptive mesh can be switched on
smoothly starting from a Lagrangean description is therefore a very useful
feature.

We have applied the code to a typical Cepheid model pulsator.  The
temporal variations of the spatial structure of the temperature are well
tracked with the adaptive mesh.  This results in much smoother velocity and
luminosity curves than can be obtained with the Lagrangean code.

\acknowledgements

We are indebted to G\'eza Kov\'acs for fruitful discussions at the beginning of
this project.  We also gratefully acknowledge the support of NSF
(AST-92-18068), and of an RCI grant by the University of Florida and by IBM at
the NERD Center.

\end{document}